\documentclass[12pt]{article}
\usepackage{latexsym}
\usepackage{graphicx,color}

\begin{document}

\title{Angular momentum effects in Michelson~-~Morley type experiments}
\author{Angelo Tartaglia \thanks{%
Prof. Angelo Tartaglia, Dipartimento di Fisica del Politecnico di Torino,
Corso Duca degli Abruzzi 24, I-10129 Torino - ITALY, tel. \ +390115647328,
fax \ +390115647399} , Matteo Luca Ruggiero \\
Dip. Fisica, Politecnico and INFN, Turin, Italy \\
e-mail: tartaglia@polito.it, ruggierom@polito.it}
\maketitle

\begin{abstract}
The effect of the angular momentum density of a gravitational source on the
times of flight of light rays in an interferometer is analyzed. The
calculation is made imagining that the interferometer is at the equator of
an axisymmetric steadily rotating gravity source. In order to evaluate the
size of the effect in the case of the Earth a weak field approximation for
the metric elements is introduced. For laboratory scales and non-geodesic
paths the correction due to the angular momentum turns out to be comparable
with the sensitivity expected in gravitational waves interferometric
detectors, whereas it drops under the threshold of detectability when using
free (geodesic) light rays.
\end{abstract}

\setcounter{footnote}{3} \renewcommand{\thefootnote}{\fnsymbol{footnote}}

\setcounter{footnote}{0}

{\small \indent PACS number 04.80.C \newline
 }{\normalsize %
\pagebreak }

\section{\protect\normalsize Introduction}

{\normalsize \setlength{\baselineskip}{25pt} The famous Michelson-Morley
experiment does not require any explanation regarding its nature and the
crucial role that history reserved to it is well known. It has been analyzed
in any respect in the early days of relativity and discussed also on its
fundamental meaning \cite{reichenbach}. Since then it has been assumed that
no anisotropy can be revealed until the frontier of special relativity is
crossed. Only in a few cases anisotropies deriving from general relativistic
corrections were considered \cite{zhang}\cite{diaz}, but only caused by the
gravitational red shift in non-horizontal arms of the interferometer;
Schwarzschild-like corrections do not produce any effect in the horizontal
plane. }

{\normalsize However, if the source of the field is rotating as it is the
case for the Earth, the situation in principle changes. This means that a
tiny anisotropy can legitimately be expected, depending on the angular
momentum of the source. }

{\normalsize On the other hand the search for measurable effects of the
angular momentum of the gravitational field is always active in order to add
a new direct verification of the consequences of general relativity. The
only positive result at the moment concerns the precession of the nodes of
the orbit of the LAGEOS satellite \cite{ciufolini} (Lense-Thirring effect
\cite{lense}). In the next few years the space mission Gravity Probe B (GPB)
is planned to fly carrying gyroscopes which should in turn verify the
Lense-Thirring effect too \cite{GPB}; finally a series of different
possibilities connected both with the Sagnac effect and the gravitomagnetic
clock effect have been considered \cite{tartaglia}\cite{mashhoon}. }

{\normalsize The present paper will fix the general formalism to verify a
possible influence of the angular momentum density of the Earth on a
Michelson-Morley type experiment. Numerical estimates will show that the
effect is quite small in any case, however using non-geodesic light paths it
could turn out to be comparable with what people are expecting and planning
to measure with big size interferometric gravitational waves detectors like
LIGO \cite{ligo1} and VIRGO \cite{virgo1}. }

\section{\protect\normalsize Preliminaries}

{\normalsize The Michelson-Morley experiment is an interferometric measure
and uses light, let us then start from a generic null line element in polar
coordinates and within an axially symmetric static field originated by a
central body endowed with an angular velocity $\Omega =d\phi /dt$:
\begin{equation}
0=g_{tt}dt^{2}+2g_{t\phi }dtd\phi +g_{rr}dr^{2}+g_{\theta \theta }d\theta
^{2}+g_{\phi \phi }d\phi ^{2}  \label{generale}
\end{equation}
The $g$'s are of course the elements of the metric and are independent both
from time $t$ and from $\phi $. In weak-field approximation the explicit
form of the metric can be \cite{straumann}
\begin{eqnarray}
g_{tt} &\simeq &c^{2}\left( 1-\frac{\mu }{r}\right)  \nonumber \\
g_{t\phi } &\simeq &\frac{\mu c}{r}a\sin ^{2}\theta  \nonumber \\
g_{rr} &\simeq &-1-\frac{\mu }{r}+\frac{\sin ^{2}\theta }{r^{2}}a^{2}
\label{weak} \\
g_{\theta \theta } &=&-r^{2}-a^{2}\cos ^{2}\theta  \nonumber \\
g_{\phi \phi } &\simeq &-\left( r^{2}+a^{2}\right) \sin ^{2}\theta  \nonumber
\end{eqnarray}
where we introduced the parameters $a=J/Mc$ ($J$ is the angular momentum of
the source, $M$ is its mass and $c$ is the speed of light), $\mu =2GM/c^{2}$
(Schwarzschild radius of the source). Now let us consider $r=$ constant
world lines only. This choice corresponds to limiting the study to light
beams contained locally in a ''horizontal'' plane (actually this would
require a wave guide locally shaped as a constant gravitational potential
surface). The null (non-geodesic) world line becomes:
\begin{equation}
0=g_{tt}dt^{2}+2g_{t\phi }dtd\phi +g_{\theta \theta }d\theta ^{2}+g_{\phi
\phi }d\phi ^{2}  \label{rcostante}
\end{equation}
If we suppose to place our interferometer at the equator ($\theta =\pi /2$)
and provided its arms are not too long, the metric (\ref{weak}) (first order
in $a/r$, $\mu /r$ and $\frac{\pi }{2}-\theta $) becomes
\begin{eqnarray}
g_{tt} &\simeq &c^{2}\left( 1-\frac{\mu }{r}\right)  \nonumber \\
g_{t\phi } &\simeq &\frac{\mu c}{r}a  \nonumber \\
g_{rr} &\simeq &-1-\frac{\mu }{r}+\frac{a^{2}}{r^{2}}  \label{weak1} \\
g_{\theta \theta } &\simeq &-r^{2}  \nonumber \\
g_{\phi \phi } &\simeq &-r^{2}-a^{2}  \nonumber
\end{eqnarray}%
\newline
Consequently for short enough excursions in the ''horizontal plane'' we can
assume, at the lowest order in $\theta $, that, for light, $\phi $ and $%
\theta $ variations are approximately proportional to each other, so:
\begin{equation}
\left| d\theta \right| =\chi \left| d\phi \right|  \label{thetaphi}
\end{equation}
where $\chi $ is a constant. }

{\normalsize Suppose now that the interferometer arms are stretched one in
the North-South direction and the other in the East-West direction. Taking
into account the fact that the Earth reference frame, where the
interferometer is at rest, is indeed rotating, the coefficient $\chi $ will
depend on the angular speed of the Earth. }

{\normalsize Now solving (\ref{rcostante}) for $dt$ we obtain
\begin{equation}
dt=\frac{-g_{t\phi }\pm \sqrt{g_{t\phi }^{2}-g_{tt}g_{\theta \theta }\chi
^{2}-g_{tt}g_{\phi \phi }}}{g_{tt}}d\phi  \label{differenziale}
\end{equation}
}

{\normalsize Supposing our light beam starts from a point on the equator at $%
\phi =0$ and moves northward, it will be
\begin{equation}
\theta =\frac{\pi }{2}-\chi \phi  \label{equatore}
\end{equation}
Of course in the case of an East-West beam it is $\chi =0$. \newline
}

\section{\protect\normalsize Times of flight of non-geodesic light beams}

{\normalsize In proximity of the equator the first factor in the right hand
side of (\ref{differenziale}) does not depend on $\theta $; thus
\begin{equation}
t_{N}=\frac{-g_{t\phi }+\sqrt{g_{t\phi }^{2}-g_{tt}g_{\theta \theta }\chi
^{2}-g_{tt}g_{\phi \phi }}}{g_{tt}}\phi _{1}  \label{uno}
\end{equation}
$t_{N}$ is the time of flight to reach the northern mirror and $\phi _{1}$
is the angular coordinate of the event; the drift of the beam is naturally
in the prograde sense. }

{\normalsize The world line of the mirror (initially at $\theta =\frac{\pi }{%
2}-\Phi $ and $\phi =0$;$\;$here $\Phi $ represents the angular stretch of
the interferometer arm) is:
\begin{equation}
t_{N}=\phi _{1}/\Omega  \label{due}
\end{equation}
$\Omega $ is of course the angular speed of the Earth. }

{\normalsize (\ref{uno}) and (\ref{due}) allow to deduce an expression for $%
\chi $:
\begin{equation}
\chi =\frac{1}{\Omega }\sqrt{-\frac{g_{tt}+2g_{t\phi }\Omega +g_{\phi \phi
}\Omega ^{2}}{g_{\theta \theta }}}  \label{chi}
\end{equation}
Actually at the lowest order in $\theta $ (\ref{chi}) does not contain $%
\theta $ itself and consequently on an $r=$ $R$ $=$ constant surface $\chi $
is a constant. }

{\normalsize From (\ref{equatore}) we see that to span the South-North
angular distance $\Phi $, one travels eastward by the angle
\begin{equation}
\phi _{1}=\frac{\Phi }{\chi }  \label{span}
\end{equation}
Returning to (\ref{uno}) and using (\ref{chi}):
\[
t_{N}=\sqrt{-\frac{g_{\theta \theta }}{g_{tt}+2g_{t\phi }\Omega +g_{\phi
\phi }\Omega ^{2}}}\Phi
\]
}

{\normalsize Considering the North-South way back to the source we see from (%
\ref{uno}) that it is
\begin{equation}
t_{S}=\frac{-g_{t\phi }+\sqrt{g_{t\phi }^{2}-g_{tt}g_{\theta \theta }\chi
_{S}^{2}-g_{tt}g_{\phi \phi }}}{g_{tt}}\phi _{2}  \label{tisud}
\end{equation}
where $\phi _{2}$ is the Earth's rotation angle between the reflection and
the arrival back at the source. It must be
\begin{equation}
t_{S}=\phi _{2}/\Omega  \label{dues}
\end{equation}
(\ref{tisud}) and (\ref{dues}) give $\chi _{S}=\chi $ as in (\ref{chi});
then it is $t_{S}=t_{N}$, $\phi _{2}=\phi _{1}.$ Finally the total time of
flight South-North-South is
\[
t_{SNS}=t_{N}+t_{S}=2\sqrt{-\frac{g_{\theta \theta }}{g_{tt}+2g_{t\phi
}\Omega +g_{\phi \phi }\Omega ^{2}}}\Phi
\]
}

{\normalsize To proceed further we recall the explicit expressions for the $%
g $'s, given in (\ref{weak1}): now, posing $\Phi =l/R$ where $l$ is the
length of the arm of the interferometer and $R$ is the radius of the Earth,
one has approximately:
\[
t_{SNS}\simeq 2\frac{l}{c}\left( 1+\frac{\mu }{2R}+\frac{R^{2}\Omega ^{2}}{%
2c^{2}}+\frac{1}{2}\frac{\Omega ^{2}a^{2}}{c^{2}}\right) -2\frac{\mu a\Omega
}{c^{2}R}l
\]
All further corrections in $\theta $, i.e. $\Phi $, are indeed quadratic and
multiply the other small terms, thus resulting much smaller than them. }

{\normalsize The next step is to consider the time of flight along the
East-West arm of the interferometer. }

{\normalsize From formula (\ref{rcostante}) with $\theta =\pi /2=$ const we
have
\begin{equation}
t_{E}=\frac{-g_{t\phi }+\sqrt{g_{t\phi }^{2}-g_{tt}g_{\phi \phi }}}{g_{tt}}%
\phi  \label{est}
\end{equation}
The world line of the eastern end mirror, assuming equal length arms, is $%
\phi =\Phi +\Omega t$ which means also
\begin{equation}
\phi _{E}=\Phi +\Omega t_{E}  \label{destra}
\end{equation}
Combining (\ref{est}) and (\ref{destra}) one has
\begin{equation}
t_{E}=-\frac{g_{t\phi }-\sqrt{\left( g_{t\phi }^{2}-g_{tt}g_{\phi \phi
}\right) }}{g_{tt}+g_{t\phi }\Omega -\sqrt{\left( g_{t\phi
}^{2}-g_{tt}g_{\phi \phi }\right) }\Omega }\Phi  \label{test}
\end{equation}
}

{\normalsize Now for the way back we have
\begin{equation}
t_{W}=\frac{g_{t\phi }+\sqrt{g_{t\phi }^{2}-g_{tt}g_{\phi \phi }}}{g_{tt}}%
\left( \phi _{E}-\phi _{W}\right)  \label{ovest}
\end{equation}
$\phi _{W}$ is the angular coordinate of the source at the arrival time of
the reflected beam. It must also be
\begin{equation}
\phi _{W}=\Omega \left( t_{E}+t_{W}\right)  \label{sorgente}
\end{equation}
}

{\normalsize (\ref{sorgente}) and (\ref{ovest}) together allow for the
calculation of $t_{W}$:
\begin{equation}
t_{W}=\frac{g_{t\phi }+\sqrt{g_{t\phi }^{2}-g_{tt}g_{\phi \phi }}}{%
g_{tt}+g_{t\phi }\Omega +\sqrt{g_{t\phi }^{2}-g_{tt}g_{\phi \phi }}\Omega }%
\Phi  \label{tovest}
\end{equation}
The total West-East-West time of flight is
\begin{equation}
t_{WEW}=t_{E}+t_{W}=2\frac{\sqrt{g_{t\phi }^{2}-g_{tt}g_{\phi \phi }}}{%
g_{tt}+2g_{t\phi }\Omega +\Omega ^{2}g_{\phi \phi }}\Phi  \label{testovest}
\end{equation}
}

{\normalsize The difference in the time of flight along the two arms at the
lowest order weak field approximation is
\begin{equation}
\Delta t=t_{WEW}-t_{SNS}\simeq \frac{a^{2}}{R^{2}}\frac{l}{c}
\label{differenza}
\end{equation}
}

\section{\protect\normalsize Geodesic light beams}

{\normalsize The situation for free, i.e. geodesic, light rays is different
from the description given in the previous section. Now we start from the
remark that in the equatorial plane the bending of the light rays is lower
than the curvature of the circle along which the mirrors of the
interferometer move. Actually in the zeroth order of approximation the time
of flight of light between the two end mirrors of an arm of the
interferometer is deduced from the length of the chord subtended to the
appropriate arc of the mirrors circumference. It is
\begin{equation}
t_{o}=2\frac{R}{c}\sin \frac{\phi _{o}}{2}  \label{tzero}
\end{equation}
where $\phi _{o}=\Phi \pm \Omega t$ ($+$ in the prograde path, $-$ in the
reverse trip). }

{\normalsize Taking into account the effect of the mass $M$ and the angular
momentum density $a$, we expect a deviation from the straight line, which
can be expressed in terms of the space curvature of the light beam $k=1/\rho
$ ($\rho $ is the radius of curvature). }

{\normalsize A further approximation can be to use the average curvature of
the path between the two ends of the interferometer arc. The length of the
intercepted beam would then be
\begin{equation}
\rho \psi  \label{arco}
\end{equation}
where $\psi $ is the angle subtending the moving interferometer arc, as seen
from the curvature center. When the space trajectory of the light rays is
not contained in the equatorial plane, we expect it also no more to be plane
at all, however reasonably the non-planarity corrections will be smaller
than the other corrections we are introducing. }

{\normalsize The chord subtended to the arc (\ref{arco}) is of course the
same when seen from the origin of the reference frame; in the equatorial
plane this gives
\[
\rho \sin \frac{\psi }{2}=R\sin \frac{\phi }{2} \label{psii}
\]
Now $\phi $ is slightly different from the former $\phi _{o}$:
\[
\phi =\phi _{o}\pm \Omega \delta t
\]
where $\delta t$ is the correction to the time of flight induced by the
curvature of the trajectory. }

{\normalsize Extracting $\psi $ from (\ref{psii}):
\[
\psi =2\arcsin \left( \frac{R}{\rho }\sin \frac{\phi }{2}\right)
\]
and the time of flight becomes
\[
t=\frac{\rho }{c}\psi =2\frac{\rho }{c}\arcsin \left( \frac{R}{\rho }\sin
\frac{\phi }{2}\right)
\]
More explicitly in terms of $\delta t$ and using (\ref{tzero}):
\[
\delta t=t-t_{o}=2\frac{\rho }{c}\arcsin \left( \frac{R}{\rho }\sin \frac{%
\phi _{o}\pm \Omega \delta t}{2}\right) -2\frac{R}{c}\sin \frac{\phi _{o}}{2}
\]
Reasonably it is $\rho >>R$ and all angles are small ($<10^{-6}$ rad, which
is the angle subtended under a $1$ m arm, from the center of the Earth).
This allows for series developments up to the lowest meaningful orders, thus
\newline
\[
\delta t=\pm \frac{R}{c}\Omega \delta t+\frac{1}{24}\frac{R^{3}}{c\rho ^{2}}%
\left( \phi _{o}^{3}\pm 3\phi _{o}^{2}\Omega \delta t\right)
\]
Solving for $\delta t$
\begin{equation}
\delta t=\frac{1}{24}\frac{l^{3}}{c\rho ^{2}}  \label{corrige}
\end{equation}
}

{\normalsize Now we need an explicit expression for $\rho $. A standard
approach \cite{straumann} moves from considering the right hand side of (\ref%
{rcostante}) divided by $d\lambda ^{2}$ (where $\lambda $ is an affine
parameter) as the Lagrangian of the light ray. The cyclicity of the $t$ and $%
\phi $ coordinates leads to the constants of the motion
\begin{eqnarray*}
E &=&g_{tt}\frac{dt}{d\lambda }+g_{t\phi }\frac{d\phi }{d\lambda } \\
L &=&g_{t\phi }\frac{dt}{d\lambda }+g_{\phi \phi }\frac{d\phi }{d\lambda }
\end{eqnarray*}
wherefrom
\begin{eqnarray*}
\frac{d\phi }{d\lambda } &=&\frac{g_{t\phi }E-g_{tt}L}{g_{t\phi
}^{2}-g_{tt}g_{\phi \phi }} \\
\frac{dt}{d\lambda } &=&\frac{-g_{\phi \phi }E+g_{t\phi }L}{g_{t\phi
}^{2}-g_{tt}g_{\phi \phi }}
\end{eqnarray*}
}

{\normalsize Using the Lagrangian in the equatorial plane one obtains also
\begin{equation}
\left( \frac{dr}{d\lambda }\right) ^{2}=\frac{2ELg_{t\phi }-E^{2}g_{\phi
\phi }-L^{2}g_{tt}}{g_{rr}\left( g_{tt}g_{\phi \phi }-g_{t\phi }^{2}\right) }
\label{errepunto}
\end{equation}
and
\begin{equation}
(r^{\prime })^{2}=\frac{g_{tt}g_{\phi \phi }-g_{t\phi }^{2}}{g_{rr}}\frac{%
2jg_{t\phi }-g_{\phi \phi }-j^{2}g_{tt}}{\left( jg_{tt}-g_{t\phi }\right)
^{2}}  \label{rprimo}
\end{equation}
}

{\normalsize A $^{\prime }$ denotes differentiation with respect to $\phi $
and $j=L/E$. Introducing the variable $u=1/r$ one has
\begin{equation}
\left( u^{\prime }\right) ^{2}=u^{4}\frac{g_{tt}(u)g_{\phi \phi
}(u)-g_{t\phi }^{2}(u)}{g_{rr}(u)}\frac{2jg_{t\phi }(u)-g_{\phi \phi
}(u)-j^{2}g_{tt}(u)}{\left( jg_{tt}(u)-g_{t\phi }(u)\right) ^{2}}
\label{uprimo}
\end{equation}
Now using (\ref{weak}) (\ref{uprimo}), in the lowest order in the
gravitational parameters, becomes
\begin{equation}
\left( u^{\prime }\right) ^{2}=\frac{1}{j^{2}}-u^{2}+\mu u^{3}+2\frac{u}{%
j^{3}}\mu a+\left( \allowbreak \frac{3}{j^{2}}-2u^{2}\right) u^{2}a^{2}
\label{uprimoappro}
\end{equation}
}

{\normalsize $\allowbreak $Differentiating with respect to $\phi $ we end up
with the differential equation
\begin{equation}
u^{\prime \prime }+u=\frac{3}{2}\mu u^{2}+\frac{1}{j^{3}}\mu a-4u^{3}a^{2}+%
\frac{3}{j^{2}}ua^{2}  \label{usecondo}
\end{equation}
}

{\normalsize Now coming to the local curvature of the rays in the equatorial
plane, it is convenient to use a Cartesian reference frame in that plane
posing
\begin{eqnarray*}
y &=&\sqrt{-g_{\phi \phi }}\cos \phi \\
x &=&\sqrt{-g_{\phi \phi }}\sin \phi
\end{eqnarray*}
In terms of $u$ (which is a function of $\phi $ along the trajectory) and
using the appropriate approximation:
\begin{eqnarray*}
y &=&\frac{1}{u}\sqrt{\left( 1+a^{2}u^{2}\right) }\cos \phi \\
x &=&\frac{1}{u}\sqrt{\left( 1+a^{2}u^{2}\right) }\sin \phi
\end{eqnarray*}
then
\[
w=\frac{dy}{dx}=\frac{u^{\prime }\cos \phi +\left( 1+u^{2}a^{2}\right) u\sin
\phi }{u^{\prime }\sin \phi -\left( 1+u^{2}a^{2}\right) u\cos \phi }
\]
Finally
\begin{equation}
\frac{1}{\rho }=\frac{dw}{dx}=\mathcal{A}\left( \phi \right) +\mathcal{B}%
\left( \phi \right) a^{2}  \label{rho}
\end{equation}
where $\mathcal{A}\left( \phi \right) $ and $\mathcal{B}\left( \phi \right) $
are rather complicated functions of $u$, $u^{\prime }$, $u^{\prime \prime }$
and $\phi $. }

{\normalsize The formula (\ref{rho}) can be explicitly written in a
convenient form if the curvature is calculated at the point of closest
approach of the ray to the center of the Earth (maximum value of $u$ which
we are calling $u_{m}$). There we expect $u^{\prime }=0$ and can decide that
$\phi =0$. Reasonably the average value of the curvature along the path of
the light differs from the value at the closest approach by small
corrections. }

{\normalsize The curvature value we shall use is then
\[
\frac{1}{\rho }=-\left( u_{m}^{\prime \prime }+u_{m}\right) +\frac{1}{2}%
\left( u_{m}+3u_{m}^{\prime \prime }\right) u^{2}a^{2}
\]
Actually the last term is 12 orders of magnitude smaller than the first, so
in practice
\begin{equation}
\frac{1}{\rho }=-\left( u_{m}^{\prime \prime }+u_{m}\right)  \label{unsurho}
\end{equation}
}

{\normalsize Now $u_{m}$ can be obtained equating (\ref{uprimoappro}) to $0$%
. At the lowest approximation order it is
\[
u_{m}=\frac{1}{j}
\]
Introducing this result into (\ref{usecondo}) and (\ref{unsurho}) produces
\begin{eqnarray*}
\frac{1}{\rho ^{2}} &=&\left( \frac{3}{2}\mu \frac{1}{j^{2}}+\frac{1}{j^{3}}%
\mu a-\frac{1}{j^{3}}a^{2}\right) ^{2} \\
&=&\allowbreak \frac{9}{4}\frac{\mu ^{2}}{j^{4}}-3\frac{\mu }{j^{5}}a^{2}
\end{eqnarray*}
Now back to the time difference (\ref{corrige}). A further crude
approximation can be to put $j\simeq R$:
\begin{equation}
\delta t\simeq \frac{1}{24}\frac{l^{3}}{c}\frac{\mu }{R^{4}}\left( \frac{9}{4%
}\mu -3\frac{a^{2}}{R}\right)  \label{auto}
\end{equation}
The correction, as we see, is negligibly small: $\sim 10^{-36}$ s for the
pure mass term and $\sim 10^{-40}$ s for the $a^{2}$ term. }

{\normalsize Out of the equatorial plane we cannot reasonably expect the
situation to be different. Of course the difference in flight times along
the two arms cannot but be less or at most equal in the order of magnitude
to (\ref{auto}). In practice this means that for free light rays in the
terrestrial environment the straight line approximation is fairly adequate
and the time of flight difference has the typical $0$ value of the
Michelson-Morley experiment. }

\section{\protect\normalsize Discussion and conclusion}

{\normalsize The result (\ref{differenza}) is obtained under the assumption
that a physical apparatus (bidimensional wave guide) obliges the light rays
to move along constant radius paths. Were this possible the order of
magnitude estimate at the surface of the Earth for $1$ m long interferometer
arms would be:
\begin{equation}
\Delta t\sim 10^{-20}\ \mathit{s}  \label{numero}
\end{equation}%
This effect is purely rotational and rather small but not entirely
negligible. Should the interferometer rotate in the horizontal plane, the
time of flight difference $\Delta t$ would alternatively change of sign
displaying an oscillating behavior which in principle could be detected. In
general (\ref{numero}) fixes the scale for }$a^{2}$ effects on the Earth.

A way to increase the value of (\ref{numero}) would be to have multiple
reflections of the light beams happen along each arm of the interferometer
before the actual interference is measured. This is what happens in
Fabry-Perot type interferometers. Here both arms should be equipped with
such devices, it would then be easy to have the light rays to be reflected
back and forth for instance $10^{3}$ or $10^{4}$ times, before the
measurement. Since the effect we are looking at is indeed cumulative, this
fact would bring the time difference between the two paths to $%
10^{-17}-10^{-16}$ s, which corresponds, for visible light, to $%
10^{-2}-10^{-1}$ fringe in the interference pattern. This would indeed be a
huge effect: $10^{-4}-10^{-2}$ fractional change of the signal intensity.

{\normalsize It is remarkable that the obtained numeric value compares with
the expected phase (and time) shifts in the gravitational wave
interferometric detectors now under construction, as LIGO and VIRGO \cite%
{ligo1},\cite{virgo1}. There indeed a sensitivity is expected, in measuring
displacements, of the order of $10^{-16}$ m which corresponds to a time of
flight difference 4 orders of magnitude lower than (\ref{numero}) }and
consequent much higher sensitivity{\normalsize . }

Of course our effect as such is a static one, producing a DC signal and it
would be practically impossible to recognize its presence in the given
static interference pattern. On the other hand the spectacular sensitivity
of {\normalsize gravitational wave interferometric detectors is obtained at
frequencies in the range }$10^{2}-10^{3}$ Hz. To extract the information
from the background and to profit of highly refined interferometric
techniques we would have to modulate the signal. This result could be
achieved, for instance, steadily rotating the whole interferometer in the
horizontal plane. This solution would however introduce a new source of
noise too, in form of vibrations. It would be better to think of a static
interferometer with a couple of rotating beams. In practice one could have
two cylindric, coaxial mirrors; the internal one should be partially
transparent. On the axis one would have a compact rotating head with a
couple of source/receivers shooting two light beams orthogonally to each
other; this result would actually be performed with an appropriate beam
splitter on the axis and a primary source sending light along the axis. The
couple of cylindrical reflecting surfaces would act as a Fabry-Perot device;
the return beams would interfere on the axis.

Of course in this configuration one would have the sought for signal,
modulated at a frequency double of the rotation frequency of the beams. Any
imperfection of the mirrors and of the whole set up would of course generate
perturbations at the same fundamental frequency as the one of the signal.
The difference between signal and rotation induced noise would be that the
signal would be peaked on the East-West direction, whereas the various
disturbances would have random orientations of their axes. Repeating many
measurement runs with different angular configurations of the interferometer
and carefully analyzing the data should allow separating the East-West
peaked signal from the rest.

Besides the rotation induced perturbations one would expect also elastic
vibration and thermal noise in the solid body of the interferometer. The
elastic vibrations can be controlled carefully designing the structure in
order to have proper frequencies not coinciding with the rotation frequency
of the beams. For a rigid configuration at the scale of the meter, proper
frequencies are easily greater than 1 kH.

As for thermal noise, it can be controlled operating at low temperature. In
any case if a phase difference can be achieved in the order of one hundredth
of a fringe or better, the signal would easily be bigger than the amplitude
of the thermal noise of an even moderately cooled device.

In order to cope with the needs both of mechanical and thermal stability,
and considering that we need also to properly guide the optical beams a best
suited material could be sapphire, whose properties make it extremely
interesting for interferometry in gravitational waves detection \cite%
{sapphire}.

The one described above is a simple scheme, showing the principle
feasibility of an experiment. A careful analysis of the technical details
would of course be needed in order to proceed further.

We can conclude that the calculations we have written in this paper fix the
order of magnitude of effects depending from the $a^{2}$ of the Earth and
show that they should be big enough to be measurable by interferometric
techniques.

{\normalsize \ \setlength{\baselineskip}{10pt} }

\end{document}